%% file: hadron2011.tex
\documentclass[a4paper,11pt]{article}

\usepackage{contribution}
\input{econfmacros}
\input{contributionmacros}

\begin{document}

% % % % % % % % % % % % % % % % % % % % % % % % % % % % % % % % % % % % % % % % %
% your proceedings
\input{contribution}

\end{document}

%% file: econfmacros.tex
\newcommand{\weblink}[2][]{%
    \ifthenelse{\equal{#1}{}}%
    {\textnormal{\url{#2}}}%
    {\textnormal{\href{#2}{#1}}}%
}

\newcommand{\acknowledgements}[1]{%
  \bigskip\bigskip
  \textsf{\textbf{\Large Acknowledgements}} \\[2ex]
  {#1}
  \bigskip
}

\def\beq{\begin{equation}}
\def\eeq#1{\label{#1}\end{equation}}
\def\eeqn{\end{equation}}

\def\beqa{\begin{eqnarray}}
\def\eeqa#1{\label{#1}\end{eqnarray}}
\def\eeqan{\end{eqnarray}}

\let\bar=\overbar

\def\Dslash{\not{\hbox{\kern-4pt $D$}}}
\def\dslash{\not{\hbox{\kern-2pt $\del$}}}

\def\msb{{\bar{\ssstyle M \kern -1pt S}}}

%% file: contributionmacros.tex
\newcommand{\contribution}[7][]{%
  \clearpage
  \thispagestyle{plain}
  \ifthenelse{\equal{#1}{}}
  {\hypersetup{pdftitle={#2}}}
  {\hypersetup{pdftitle={#1}}}
  \hypersetup{pdfauthor={{#3} {#4}}}
  {\centering\normalfont\LARGE\bfseries\sffamily #2 \par\nobreak}
  \lhead{}
  \chead{%
    \textit{\footnotesize XIV International Conference on Hadron Spectroscopy
      (\weblink[\textit{hadron2011}]{http://www.hadron2011.de}), 13-17 June 2011, Munich, Germany}%
  }
  \rhead{}
  \bigskip
  \begin{center}
    {#3} {#4}\ifthenelse{\equal{#6}{}}{}{\footnote{\weblink[#6]{mailto:#6}}}
    \ifthenelse{\equal{#7}{}}{}{#7} \\
    \textit{#5}
  \end{center}
  \bigskip
}

\renewcommand{\abstract}[1]{%
  \begin{center}
    \begin{minipage}{0.85\textwidth}
      \begin{footnotesize}
        #1
      \end{footnotesize}
    \end{minipage}
  \end{center}
  \bigskip
}

%% file: contribution.tex
%%%%%%%%%%%%%%%%%%%%%%%%%%%%%%%%%%%%%%%%%%%%%%%%%%%%%%%%%%%%%%%%%%%%%%%%%%%%%%%%%
%
% template for hadron2011 contribution
%
% please do not rename this file
%
% to create document run
%
%     pdflatex hadron2011.tex
%
%%%%%%%%%%%%%%%%%%%%%%%%%%%%%%%%%%%%%%%%%%%%%%%%%%%%%%%%%%%%%%%%%%%%%%%%%%%%%%%%%
{  % do not remove

%%%%%%%%%%%%%%%%%%%%%%%%%%%%%%%%%%%%%%%%%%%%%%%%%%%%%%%%%%%%%%%%%%%%%%%%%%%%%%%%%
% please define your macros here

%
%%%%%%%%%%%%%%%%%%%%%%%%%%%%%%%%%%%%%%%%%%%%%%%%%%%%%%%%%%%%%%%%%%%%%%%%%%%%%%%%%

%%%%%%%%%%%%%%%%%%%%%%%%%%%%%%%%%%%%%%%%%%%%%%%%%%%%%%%%%%%%%%%%%%%%%%%%%%%%%%%%%
% define title, author, and address
% contribution[short title]{title}{author first name}{author last name}{author address}{author email}{collaboration}
% the short title will appear in the page headers and the TOC of the book of proceedings
% the last two arguments may be left empty
\contribution[Diffractive production of $(K\bar{K}\pi)^{0}$]% short title (optional)
{Diffractive dissociation into $\mathbold{K_sK^{\pm}\pi^{\mp}\pi^{-}}$ final states}
{Johannes~Bernhard$^{a}$~and} {Frank~Nerling$^{b,}$}  % first and last name of author
{
$^{a}$ Institut f\"ur Kernphysik, Universit\"at Mainz, 55099 Mainz, GERMANY
\\
$^{b}$ Physikalisches Institut, Albert-Ludwigs-Universit\"at Freiburg 
79104 Freiburg, GERMANY  % author address
} 
{Correspondence: nerling@cern.ch}  % author email optional
{on behalf of the COMPASS Collaboration}  % collaboration (optional)
%
%%%%%%%%%%%%%%%%%%%%%%%%%%%%%%%%%%%%%%%%%%%%%%%%%%%%%%%%%%%%%%%%%%%%%%%%%%%%%%%%%

%%%%%%%%%%%%%%%%%%%%%%%%%%%%%%%%%%%%%%%%%%%%%%%%%%%%%%%%%%%%%%%%%%%%%%%%%%%%%%%%%
% abstract
\abstract{%
The COMPASS fixed-target experiment at CERN/SPS is dedicated to the study of hadron
structure and spectroscopy, especially the search for spin-exotic states.
% and glueballs. 
%% COMPASS has started to contribute to the puzzle of the existence of spin-exotic 
%% mesons by the published result on the 2004 data of an exotic $J^{PC}=1^{-+}$ state 
%% at 1.66\,GeV/${\rm c^2}$,
%% %\,\cite{Alekseev:2010}
%% the newly taken 2008/09 data will further clarify the situation. 
After having started to study the existence of the spin-exotic $\pi_1(1600)$ 
resonance in the 2004 pilot-run data, the new 2008/09 data will enable us to further 
clarify the situation. Apart from the $\pi_1(1600)$ resonance, 
%FN observed in various decay channels and experiments, 
also a spin-exotic $\pi_1(2000)$ was reported in 
the past in the $f_1(1285)\pi$ decay channel by the E852/BNL experiment,
%\,\cite{JKuhn:2004}
however, this state still lacks confirmation.
% by a second experiments.
We present a first event selection of the diffractively produced $(K\bar{K}\pi\pi)^{-}$ 
system showing clean $f_1(1285)$ and $f_1(1420)$ resonances at competing statistics.
%FN , extending the spectrum beyond 2\,GeV/${\rm c^{2}}$. 
A partial-wave analysis 
started on $f_1(1285)\pi$ and $f_1(1420)\pi$ decay channels will further complete the 
search for spin-exotics in the 2008/09 COMPASS data.
%, the $f_1(1420)\pi$ system was never studied before.
}
%
%%%%%%%%%%%%%%%%%%%%%%%%%%%%%%%%%%%%%%%%%%%%%%%%%%%%%%%%%%%%%%%%%%%%%%%%%%%%%%%%%

%%%%%%%%%%%%%%%%%%%%%%%%%%%%%%%%%%%%%%%%%%%%%%%%%%%%%%%%%%%%%%%%%%%%%%%%%%%%%%%%%
% main text
% for short contributions sections are optional
\vspace{-0.5cm}
\section{Introduction}
\vspace{-0.4cm}
One important part of the COMPASS physics programme is the search for new states, 
in particular the search for spin-exotic states and glueballs. 
COMPASS has started to contribute to the puzzle of the existence of spin-exotic 
mesons with the published 2004 pilot-run data, showing a significant production 
strength for an exotic $J^{PC}=1^{-+}$ state at 1.66\,GeV/$c^2$~\cite{Alekseev:2010}.
The high-statistics 2008/09 data sets, covering almost all decay channels reported in the past, 
will further clarify the situation. 
Apart from the spin-exotic $\pi_1(1600)$ resonance observed in various decay channels 
and experiments, also a spin-exotic $\pi_1(2000)$ candidate was reported in the past by the E852 
experiment at BNL in the $f_1(1285)\pi$ decay channel~\cite{JHLee:1994,JKuhn:2004}. This state, 
however, still requires confirmation.
 
In this paper, we present the event selection of the diffractively produced $(K\bar{K}\pi\pi)^{-}$ 
system showing clean $f_1(1285)$ and $f_1(1420)$ signals competitive with the BNL/E852 data. 
The accessible mass range is extended beyond 2\,GeV/$c^{2}$.
The started partial-wave analysis (PWA) of the $f_1(1285)\pi$ and $f_1(1420)\pi$ systems will 
further complete the search for spin-exotics in the 2008/09 COMPASS data. 
%FN The $f_1(1420)\pi$ system was never studied before.
A PWA of the $(K\bar{K}\pi\pi)^{-}$ system will on the one hand complement previous searches in the 
$f_1\pi$-channel, on the other hand it will be the first PWA of the $f_1(1420)\pi$ system.
%FN Moreover, it will be the first PWA of the $f_1(1420)\pi$ system.
\begin{figure}[tp!]
\vspace{-0.5cm}
  \begin{minipage}[h]{.62\textwidth}
    \begin{center}
     \includegraphics[clip, trim= 50 65 60 90,width=0.9\linewidth]{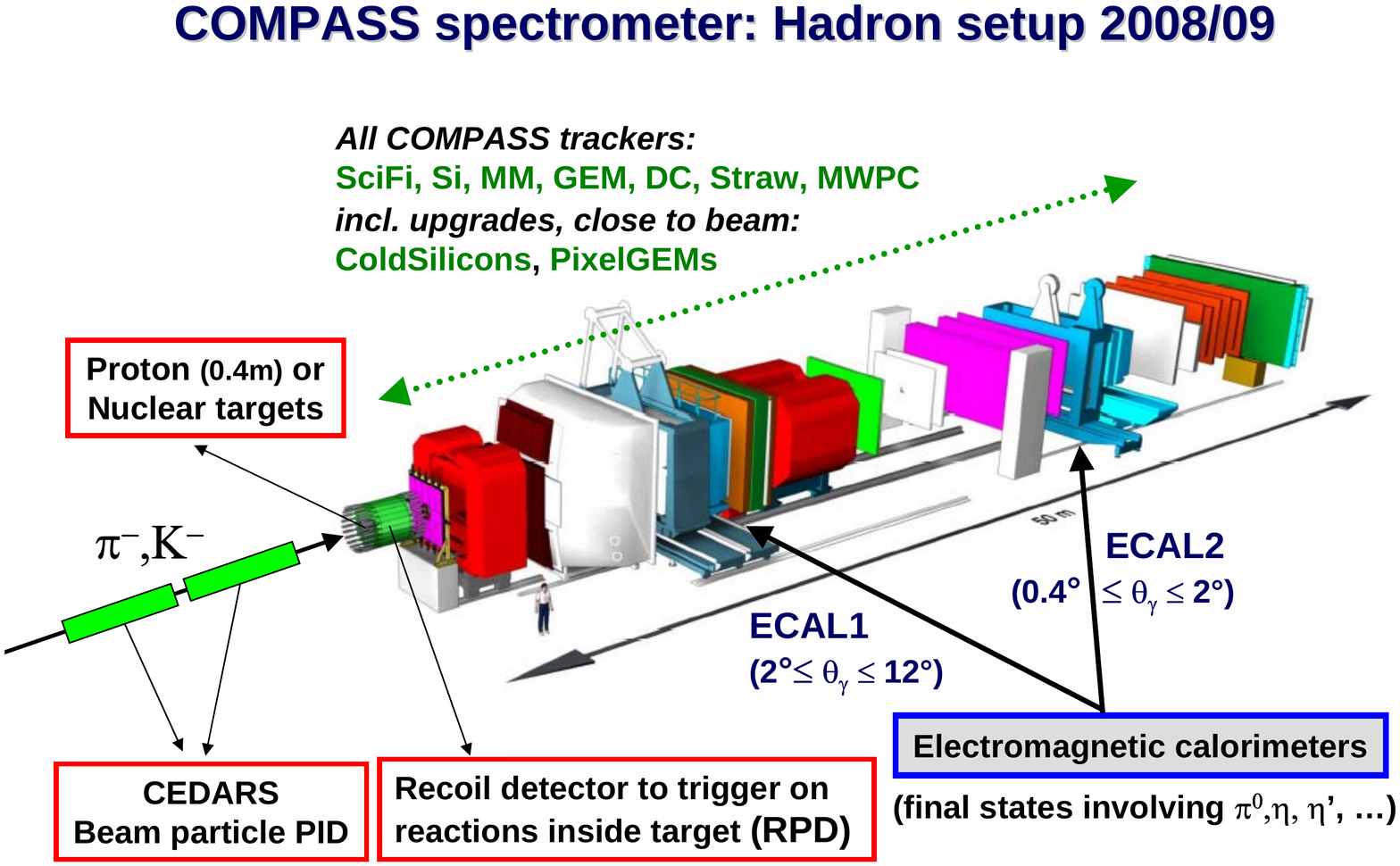}
    \end{center}
  \end{minipage}
  \hfill
  \begin{minipage}[h]{.38\textwidth}
    \begin{center}
     \includegraphics[clip, trim= 5 5 45 30,width=1.0\linewidth]{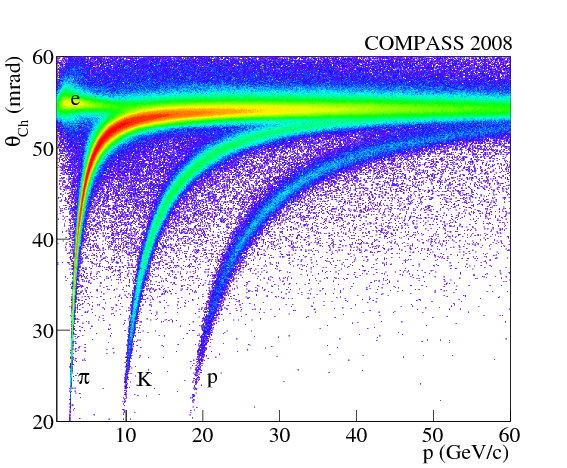}
    \end{center}
  \end{minipage}
\vspace{-0.2cm}
      \caption{{\it (Left)} Sketch of the two-stage COMPASS spectrometer.
               {\it (Right)} Measured Cherenkov angle using 
	RICH-1 versus particle momentum. Three bands appear corresponding to the different masses of
	pions, kaons, and (anti-)protons; some additional contribution from $\delta$-electrons is present at low masses and angles.}
       \label{fig:diffrProd_Spectro} 
\end{figure}

The COMPASS two-stage spectrometer~\cite{compass:2007} at the CERN SPS features charged 
particle tracking and good coverage by electromagnetic calorimetry for both stages 
(Fig.\,\ref{fig:diffrProd_Spectro}). 
%For the data presented here, a fixed liquid hydrogen target surrounded by a 
%recoil proton detector (RPD) included in the trigger was used.   
The fixed liquid hydrogen target is surrounded by a recoil proton detector (RPD) included 
in the trigger. Moreover, a Ring Imaging Cherenkov (RICH) detector in 
the first stage allows for final state particle identification (PID). 
A good separation of pions from kaons allows for final state PID and hence the study of kaonic final states.
%FN enables the study of kaonic final states. 
Two Cherenkov Differential counters with Achromatic Ring focus (CEDAR) upstream of the 
target are used to identify the incoming beam particle.  
Not only kaon diffraction, tagging the kaon contribution in the negative hadron beam 
(96.8\,\% $\pi^{-}$, 2.4\,\% $K^{-}$, 0.8\,\% $\bar{p}$) can thus be studied, cf.~\cite{PJasinski:2010}, 
but also production of strangeness with the pion beam, see also~\cite{tobi:2009,nerling:2010}. 
%FN Apart of the $(K\bar{K}\pi)^{0}$ system discussed here, the $(K\bar{K}\pi)^{-}$ system has been started to be studied 
%FN as well, cf.~\cite{tobi:2009}.
The COMPASS data recorded with 190\,GeV/$c$ hadron beams in 2008/09 provide excellent 
opportunity for simultaneous observation of new states in various decay modes within 
the same experiment, see also~\cite{nerling:2009,haas:2011,tobi:2011}. 
%% Moreover, the data contain subsets with different beam projectiles ($\pi^{\pm},K^{\pm},p$) and targets 
%% (H$_2$, Ni, W, and Pb), allowing for systematic studies not only of diffractive and central production but also Primakoff 
%% reactions~\cite{grabmueller:2010} and baryon spectroscopy~\cite{austregesilo:2010}.  

\vspace{-0.5cm}
\section{First glimpse on the diffractively produced $\mathbold{(K\bar{K}\pi)^{0}}$ system} 
\label{sec.PWAresults}
\vspace{-0.5cm}
The $(K\bar{K}\pi\pi)^{-}$ events are selected from the full 2008 data set taken with negatively charged pion beam. 
%%%%%%%%%%%%%%%%%%%%%%%%%%%%%%%%%%%%%%%%%%%%%%%%%%
\begin{figure}[tp]
\vspace{-0.5cm}
  \begin{minipage}[h]{.49\textwidth}
    \begin{center}
      \includegraphics[clip, trim= 12 10 12 20, width=1.0\textwidth]{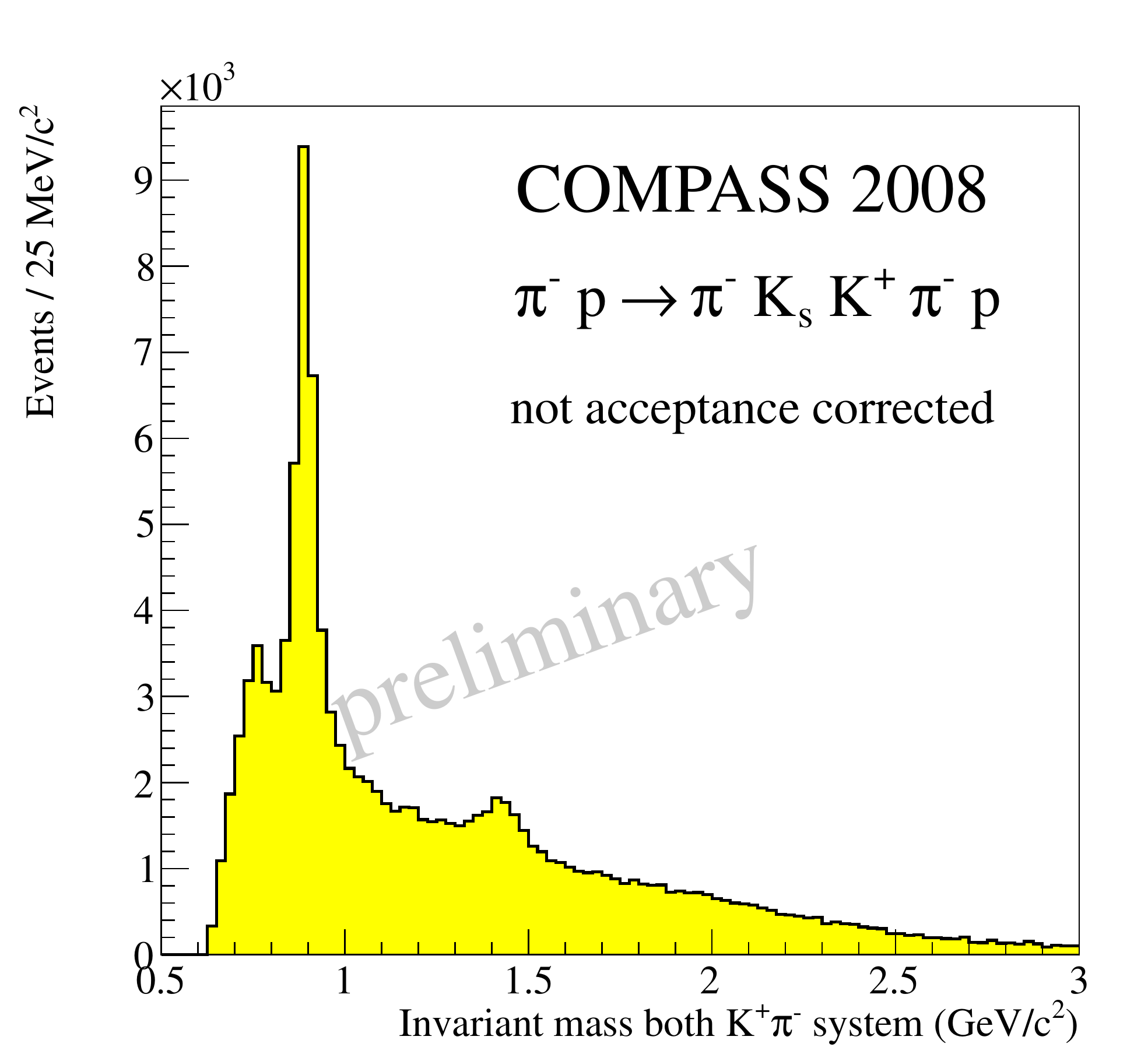}
    \end{center}
  \end{minipage}
  \hfill
  \begin{minipage}[h]{.49\textwidth}
    \begin{center}
      \includegraphics[clip, trim= 12 10 12 20, width=1.0\textwidth]{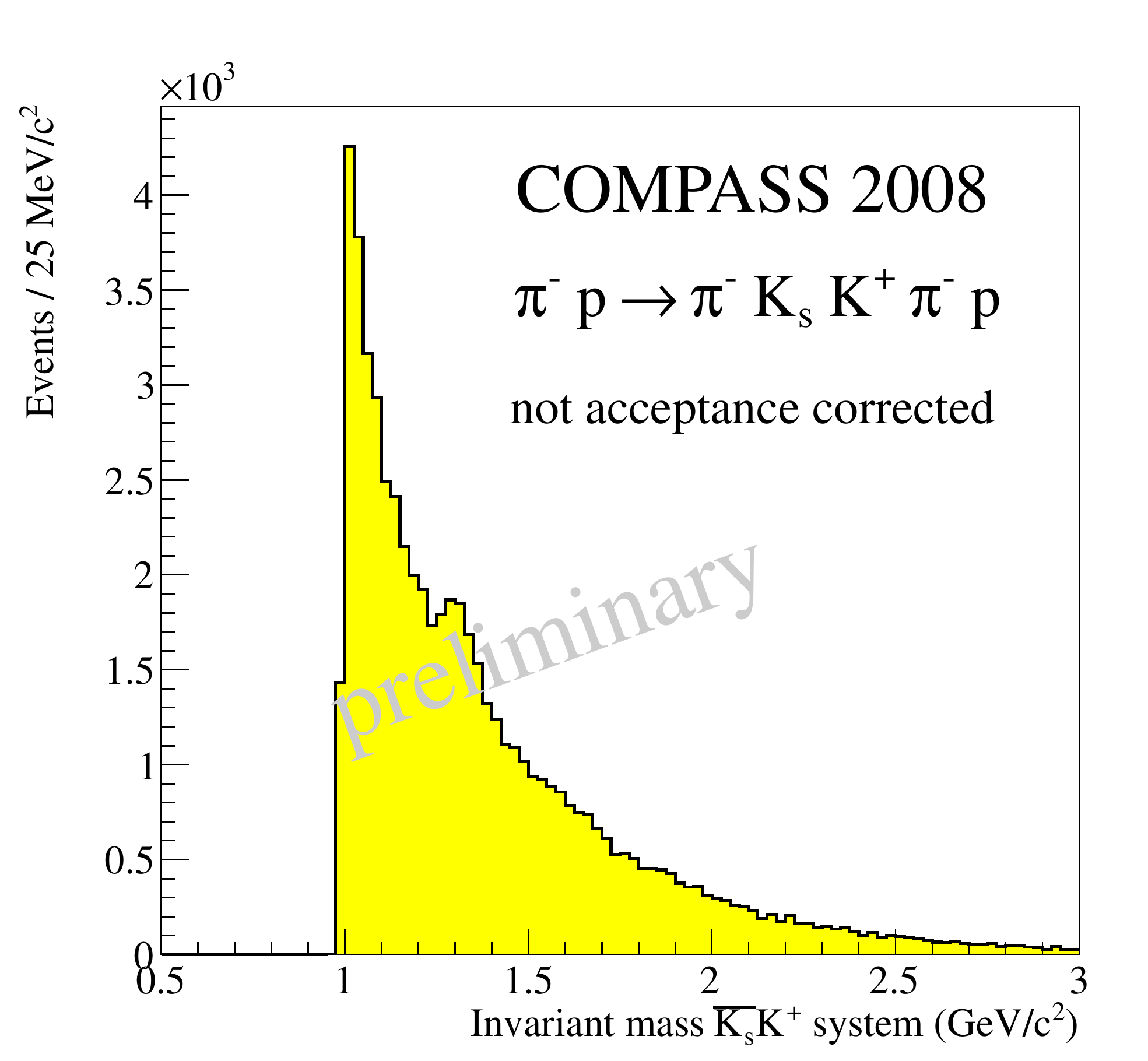}
    \end{center}
  \end{minipage}
  \begin{center}
\vspace{-0.3cm}
    \caption{Exemplary sub mass spectra shown for event type (a) (similar for (b)).
{\it Left:} The $(K\pi)^{0}$ subsystem features a clean $K^{*}(892)$ peak and some contribution from
 $K_{0,2}^{*}(1430)$.
{\it Right:} The $(K\bar{K})$ subsystem shows a clean $a_0(980)$ peak at threshold and some 
$a_2(1320)$ contribution.}
\label{fig:a0_kstar}
  \end{center}
\end{figure}
Exactly one reconstructed primary vertex inside the target volume is required for each event. 
Further, three outgoing charged tracks (two negative, one positive) were required, resulting 
in two types of final states: {\it (a)} $K_sK^{+}\pi^{-}\pi^{-}$ and {\it (b)} $K_sK^{-}\pi^{+}\pi^{-}$.
\begin{figure}[bp]
\vspace{-0.5cm}
  \begin{minipage}[h]{.49\textwidth}
    \begin{center}
      \includegraphics[clip, trim= 12 10 12 20, width=1.0\textwidth]{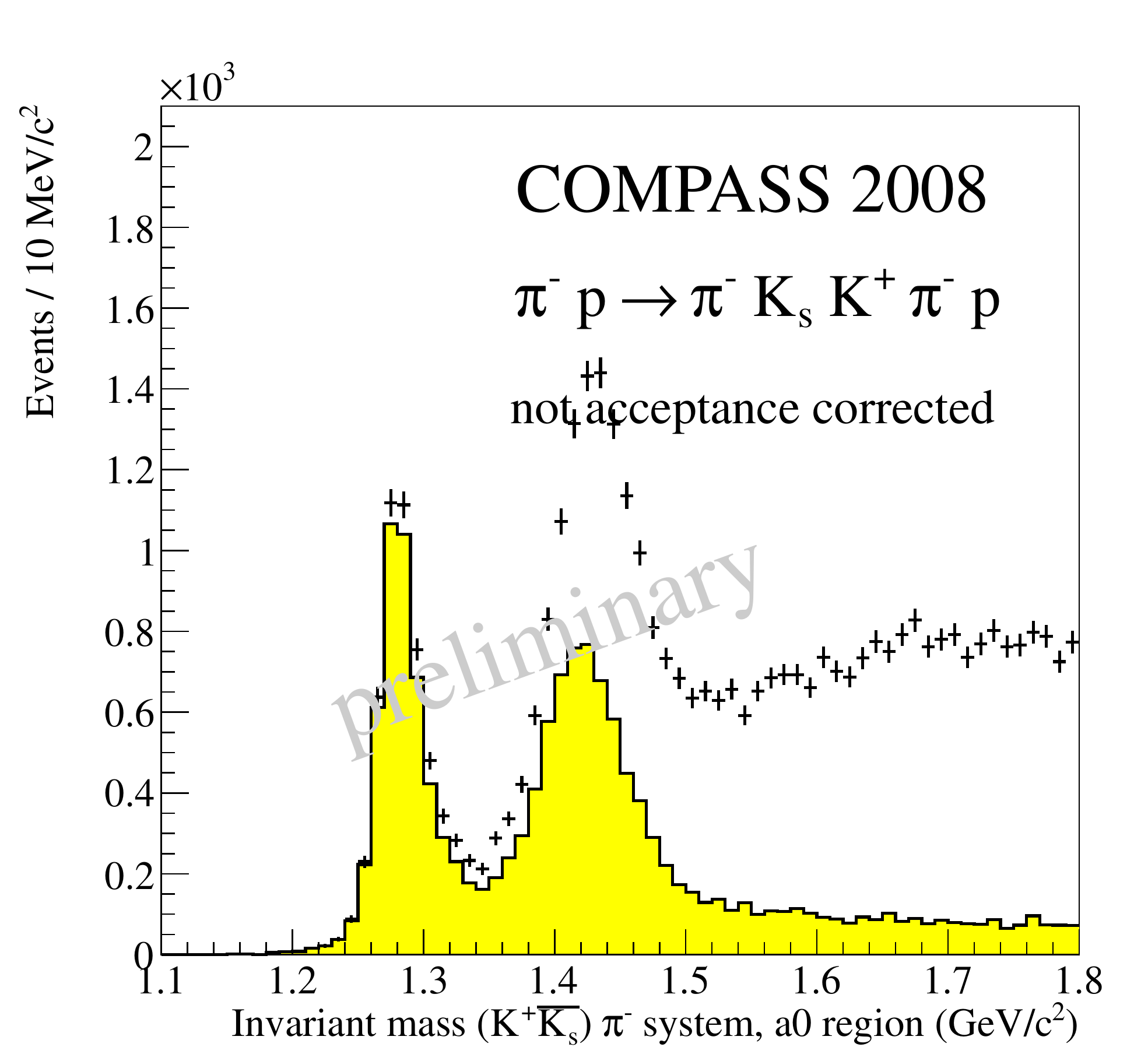}
    \end{center}
  \end{minipage}
  \hfill
  \begin{minipage}[h]{.49\textwidth}
    \begin{center}
      \includegraphics[trim = 155mm 25mm 5mm 55mm, clip, width=1.0\textwidth]{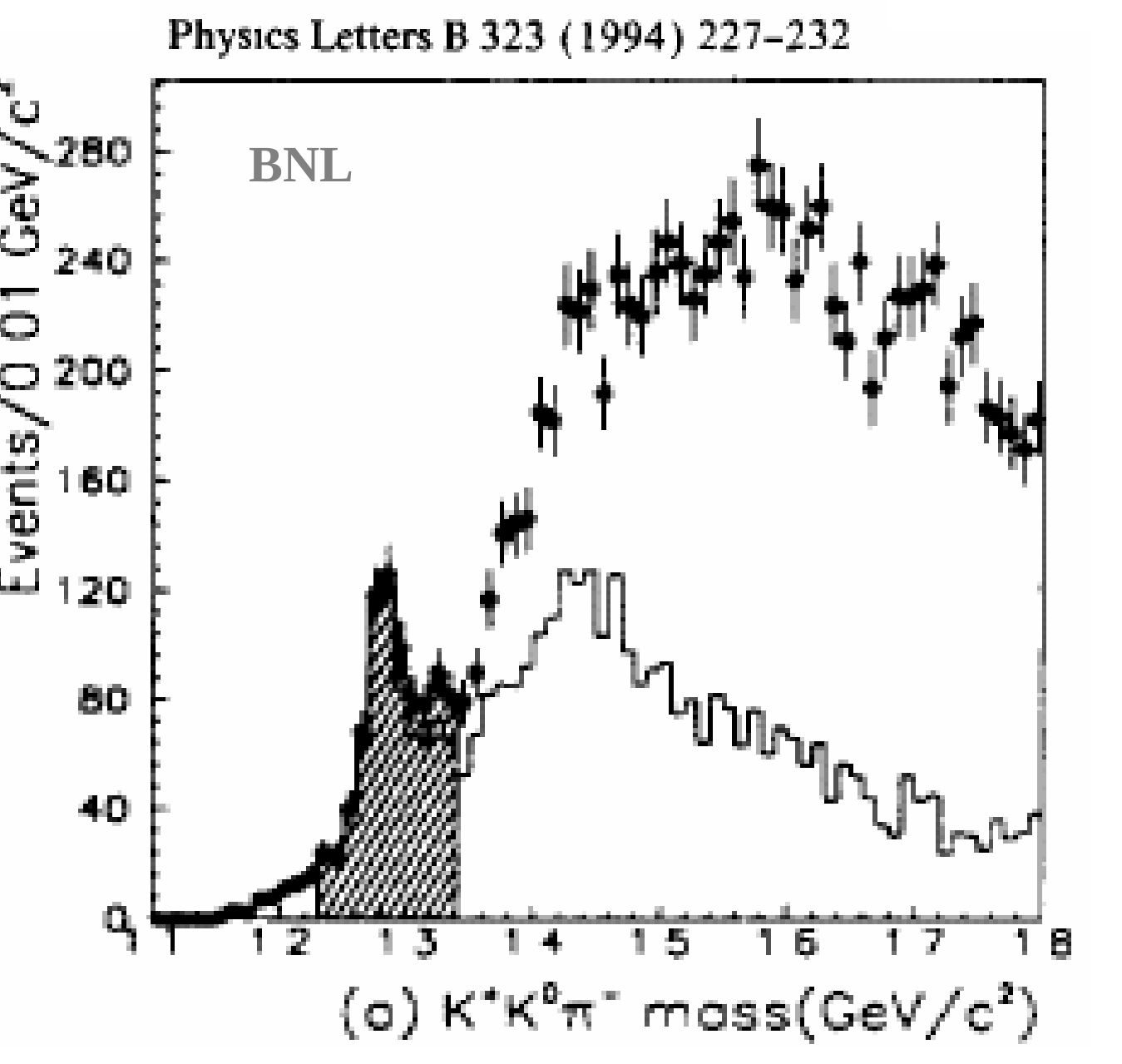}
%FN      \includegraphics[trim = 150mm 36mm 0mm 50mm, clip, width=0.90\textwidth]{Plots/techReleaseKKpipi_KKpi.pdf}
    \end{center}
  \end{minipage}
  \begin{center}
\vspace{-0.3cm}
    \caption{{\it Left:} The $(K\bar{K}\pi)^{0}$ subsystem, showing clean $f_1(1285)$ and $f_1(1420)$ peaks 
before (dots) and after (line) an additional restriction of the $K\bar{K}$ mass to the $a_0(980)$ region. 
{\it Right:} Comparing the similar plot published by BNL/E852~\cite{JHLee:1994}, the COMPASS statistics exceeds 
the one analysed by E852 by a factor of 10 (or a factor of about 20, taking into account also the 
2009 data with negative pion beam). Not only the observed $f_1(1285)$ but also the $f_1(1420)$ are nearly 
background free as compared to BNL/E852~\cite{JHLee:1994}.}
\label{fig:BNL}
  \end{center}
\end{figure}
The $K^{0}$s are identified by requiring the invariant mass of the pion pair from the reconstructed $V^0$ 
decay (after anticut on $\Lambda$ and $\bar{\Lambda}$) to be within $\pm 20$\,MeV/$c^{2}$ with respect to the PDG mass. 
%FN A minimum angle $\theta$ is required between the SV of the $K^{0}$ candidate ($\cos(\theta)< 0.999$). 
Only events with exactly one $K^{0}$ candidate are accepted.
By PID of the $K^{\pm}$ with the RICH detector (using the likelihood ratio of the Kaon assumption over others), 
the two types are separated, remaining particles are assumed to be pions. 
The beam energy is determined by the total momentum of the outgoing system. 
In order to select exclusive events, two main cuts are applied consistently in terms of 
$\pm 2\,\sigma$ of each distribution: On the calculated beam energy ($\pm 2$\,GeV$/c^{2}$), 
and on the azimuthal angle anti-correlation ($\pm 0.3$\,rad) of the recoil proton 
(measured with the RPD) and the outgoing system (measured with the spectrometer). 
%\clearpage
To get a first glimpse on the isobars to be included for the PWA, the mass spectra of the different 
subsystems have been studied. In Fig.\,\ref{fig:a0_kstar}, the $(K\pi)^{0}$ and the $(K\bar{K})$ 
subsystems are exemplary shown, featuring clean $K^{*}(892)$ and $a_0(980)$ peaks, also contributions
from $K_{0,2}^{*}(1430)$ and $a_2(1320)$ are present. 
%%%%%%%%%%%%%%%%%%%%%%%%%%%%%%%%%%%%%%%%%%%%%%%%%%
The $(K\bar{K}\pi)^{0}$ subsystem (Fig.\,\ref{fig:BNL}) is of particular interest, as a spin-exotic 
$1^{-+}$ resonance was reported in the $f_1(1285)\pi$ decay channel. It features clean $f_1(1285)$ and 
$f_1(1420)$ peaks. Even though an $\eta$ contribution cannot be excluded, a first mass-independent PWA 
indicate contributions from $\eta(1405)$ and $\eta(1295)$ to be minor, consistent with the observation 
by E852\cite{JKuhn:2004}.
Further isobars obviously to be included are:
%FN $K_{0}^{*}(800)$ or $\kappa$ 
$K\pi$ and $KK$ s-wave contributions and, for event type {\it (b)} only,
 $\rho(770)$, $f_1(1270)$, and $f_0(600)$ or $\sigma$ (corresponding mass spectra not shown).   

\vspace{-0.4cm}
\section{Conclusions \& summary}
\vspace{-0.3cm}
The COMPASS hadron data taken in 2008/09 will allow us to contribute solving the puzzle of light 
spin-exotic mesons, even extending the region to higher masses beyond 2\,GeV/$c^2$. 
We presented a first selection of $K_sK^{\pm}\pi^{\mp}\pi^{-}$ events, verifying
our feasibility to study not only the diffractively produced $f_1(1285)\pi$ at competing 
statistics but also the $f_1(1420)\pi$ system. In the former decay mode, apart from the $\pi_1(1600)$, 
a second spin-exotic $J^{PC}=1^{-+}$ resonance at 2\,GeV/$c^{2}$, the $\pi_1(2000)$, was reported in the past 
by one experiment~\cite{JKuhn:2004}, whereas the latter has never been studied before.   
%% The competing statistics and the possibility of detecting final states involving 
%% neutral particles as well as kaonic final states allow for simultaneous observation and confirmation 
%% of new states within the same experiment in different decay channels. 
%% \\
%% KKpipi -- f1pi channel -- pi1(2000), .   
%%%%%%%%%%%%%%%%%%%%%%%%%%%%%%%%%%%%%%%%%%%%%%%%%%%%%%%%%%%%%%%%%%%%%%%%%%%%%%%%%
% acknowledgements (optional)

\vspace{-0.3cm}
\acknowledgements{
%\vspace{-0.3cm}
This work is supported by the BMBF (Germany), especially via the ``Nutzungsinitiative CERN''.
}

%%%%%%%%%%%%%%%%%%%%%%%%%%%%%%%%%%%%%%%%%%%%%%%%%%%%%%%%%%%%%%%%%%%%%%%%%%%%%%%%%
% bibliographic items can be constructed using the LaTeX format in SPIRES
% see http://www.slac.stanford.edu/spires/hep/latex.html
% SPIRES will also supply the CITATION line information; please include it

%
%%%%%%%%%%%%%%%%%%%%%%%%%%%%%%%%%%%%%%%%%%%%%%%%%%%%%%%%%%%%%%%%%%%%%%%%%%%%%%%%%

}  % do not remove

%%% Local Variables: 
%%% mode: latex
%%% TeX-master: "../hadron2011.tex"
%%% End: 

%% file: hadron2011.bbl
\vspace{-0.7cm}
\begin{thebibliography}{99}
\bibitem{Alekseev:2010} M.~Alekseev {\it et al.}, COMPASS collaboration, {\it Phys. Rev. Lett.}, {\bf 104} (2010) {241803}.
\bibitem{JHLee:1994} J.H.~Lee {\it et al.}, {\it Phys. Lett. B} {\bf 323} (1994) {227}.
\bibitem{JKuhn:2004} J.~Kuhn {\it et al.}, {\it Phys. Lett. B} {\bf 595} (2004) {109}.
\bibitem{compass:2007} P.~Abbon {\it et al.}, COMPASS collaboration, {\it Nucl. Instrum. Meth. A} {\bf 577} (2007) {455}.
\bibitem{PJasinski:2010} P.~Jasinski, {\it These proceedings} (2011).
\bibitem{tobi:2009} T.~Schl\"uter, {\it AIP Conf. Proc.} {\bf 1257} (2010) 462.
\bibitem{nerling:2010} F.~Nerling, {\it  PoS ICHEP2010} (2010) 163; arXiv:1012.4993[hep-ex]
\bibitem{nerling:2009} F.~Nerling, {\it AIP Conf. Proc.} {\bf 1257} (2010) 286; arXiv:1007.2951[hep-ex], 
\\also {\it these proceedings} (2011).
\bibitem{haas:2011} F.~Haas, {\it These proceedings} (2011).
\bibitem{tobi:2011} T.~Schl\"uter, {\it These proceedings} (2011); arXiv:1108.6191[hep-ex].
%
%% \bibitem{Jaffe:1976} R.~Jaffe and K.~Johnsons, {\it Phys. Lett. B} {\bf 60} (1976) {201}.
%% \bibitem{Barnes:1983} T.~Barnes {\it et. al.}, {\it Nucl. Phys. B} {\bf 224} (1983) {241}.
%% \bibitem{Morningstar:2004} K.J.~Juge, J.~Kuti, C.~Morningstar, {\it AIP Conf. Proc.} {\bf 688} (2004) {193}.
%% \bibitem{MeyerHaarlem2010} C.A.~Meyer and Y.Van.~Haarlem, {\it Phys. Rev. C} {\bf 82} (2010) {025208}.
%% \bibitem{Adams:1998} G.~S.~Adams {\it et al.}, {\it Phys. Rev. Lett.} {\bf 81}, (1998) 5760.
%% \bibitem{Khokhlov:2000} Y.~Khokhlov, {\it Nucl. Phys.} {\bf A663} (2000) 596.
%% \bibitem{Amelin:2005} D.~V.~Amelin {\it et al.}, {\it Phys. Atom. Nucl.} {\bf 68} (2005) 359.
%% \bibitem{Dzierba:2006} A.R.~Dzierba {\it et al.}, {\it Phys. Rev. D} {\bf 73} (2006) {072001}.
%% \bibitem{PDG} K.~Nakamura {\it et al.} (Particle Data Group), {JPG} {\bf 37} (2010) {075021}.
%% \bibitem{Alekseev:2009a} M.~Alekseev {\it et al.}, COMPASS collaboration,   
%%   Observation of a $J^{PC}=1^{-+}$ exotic resonance in diffractive 
%%   dissociation of 190~GeV/c $\pi^-$ into $\pi^-\pi^-\pi^+$, {\it Phys. Rev. Lett}, {\bf 104} (2010) {241803}.
%% \bibitem{nerling:2009} F.~Nerling, {\it AIP Conf. Proc.} {\bf 1257} (2010) 286; arXiv:1007.2951[hep-ex].
%% \bibitem{haas:2011} F.~Haas, {\it These proceedings} (2011).
\end{thebibliography}
